# Polylogarithmic representation of radiative and thermodynamic properties of thermal radiation in a given spectral range: II. Real-body radiation

Anatoliy I Fisenko, Vladimir Lemberg

*ONCFEC Inc., 250 Lake Street, Suite 909, St. Catharines, Ontario L2R 5Z4, Canada*

*E-mail: afisenko@oncfec.com*

**Abstract**

The general analytical expressions for the thermal radiative and thermodynamic properties of a real-body are obtained in a finite range of frequencies at different temperatures. The frequency dependence of the spectral emissivity is represented as a power series. Stefan-Boltzmann's law, total energy density, number density of photons, Helmholtz free energy density, internal energy density, enthalpy density, entropy density, heat capacity at constant volume, pressure, and total emissivity are expressed in terms of the polylogarithm functions. The general expressions for the thermal radiative and thermodynamic functions are applied for the study of thermal radiation of liquid and solid zirconium carbide. These functions are calculated using experimental data for the frequency dependence of the normal spectral emissivity in the visible-near infrared range at the melting (freezing) point. The gaps between the thermal radiative and thermodynamic functions of liquid and solid zirconium carbide are observed. The general analytical expressions obtained can easily be presented in wavenumber domain.

## 1 Introduction

It is well-known that a knowledge of the spectral emissivity is necessary to measure the true temperature of a real-body using non-contact optical devices [1-3]. Therefore, a great number of experimental studies have been focused on the measurement of spectral emissivity $\varepsilon(\nu,T)$ for various materials [4-19].

Multiwavelength emissivity models to determine the surface temperature of a real-body were proposed in [20, 21]. Two of the most important emissivity models are the following: a) linear emissivity model (LEM) [22-26]; and b) log-linear emissivity model (LLE) [27-30]. The true temperature of a real-body can be measured using optical multispectral radiation thermometers in conjunction with a multiwavelength emissivity model. There are other emissivity models that base on fundamental physical principals. Such models are Maxwell, Hagen-Ruben and Edwards [31, 32].

A non-contact method for the determination of the true temperature of a real-body from "generalized" Wien's displacement law was proposed in [33-35]. The method was proven on the spectra of the thermal radiation of tungsten, tantalum and luminous flames. The accuracy in the determination of the steady-state temperature does not fall below 2% in these cases.

It is important to note that a knowledge of the frequency dependence of the spectral emissivity also allows to determine the thermal radiative and thermodynamic properties of a real-body within a finite range of frequencies. In [36-38], the thermal radiative and thermodynamic properties of some materials have been studied using spectral emissivity data presented in tabular form. These materials are: a) the hafnium, zirconium, and titanium carbides; b) $ZrB_2$-SiC-based ultra-high temperature ceramics; and c) molybdenum. The Helmholtz free energy density, internal energy density, enthalpy density, entropy density, heat capacity at constant volume, pressure, and the total emissivity were calculated numerically.

In [39], it was pointed out that there are several classes of materials and space objects, for which the thermal radiative and thermodynamic properties within a finite spectral range of frequencies can be described using the polylogarithms functions. This means that the frequency dependence of $\varepsilon(\nu,T)$ must be represented as a polynomial, for example. Some of these materials and space objects are: a) zirconium, uranium, and plutonium carbides at their melting (freezing) points [40, 41]; b) noble metals at the melting temperatures [42]; c) Fe, Co, and Ni at the melting points [43]; d) Milky Way and other galaxies [44, 45]; etc. It is essential to note that the thermal radiative and thermodynamic properties of these real-bodies with continuous spectra emitted in a finite spectral range of frequencies can be calculated analytically.

In this paper, the general analytical expressions for the thermal radiative and thermodynamic functions of a real-body are obtained using the frequency dependence of the spectral emissivity in the form $\varepsilon(\nu,T)=\sum_{i=-3}^{\infty}a_i(T)\nu^i$ . The expressions for the Stefan-Boltzmann law, total energy density, number density of photons, Helmholtz free energy density, internal energy density, enthalpy density, entropy density, heat capacity at constant volume, total emissivity, and pressure in a finite spectral range of frequencies are expressed in terms of the polylogarithm functions. This polylogarithmic representation allows us to calculate the thermal radiative and thermodynamic properties of a real-body analytically.

As an example, a study of the thermal radiative and thermodynamic properties of solid and liquid zirconium carbide is performed in detail. These properties are calculated using experimental data for the normal spectral emissivity in the spectral range $0.550\mu\mathrm{m}\le\lambda\le 0.900\mu\mathrm{m}$ at melting (freezing) temperature $T$ = 3155 K.

## 2. General relationships for thermal radiative and thermodynamic properties of a real body

The radiant spectral density of a real-body radiation can be presented in the form

$$I(\nu,T)=\varepsilon(\nu,T)I^{\mathrm{P}}(\nu,T), \tag{1}$$

where $\varepsilon(\nu,T)$ is the spectral emissivity and $I^{\mathrm{P}}(\nu,T)$ at temperature $T$ is given by the Planck law [46]:

$$I^{\mathrm{P}}(\nu,T)=\frac{8\pi h}{c^3}\frac{\nu^3}{e^{\frac{h\nu}{k_{\mathrm{B}}T}}-1} \quad . \tag{2}$$

Using the expression for the polylogarithm function of zero order [47] ($\mathrm{Li}_0(x)=\frac{x}{1-x}, |x|<1$), Eq. (2) can be written as:

$$I^{\mathrm{P}}(\nu,T)=\frac{8\pi h\nu^3}{c^3}\mathrm{Li}_0(e^{-\frac{h\nu}{k_{\mathrm{B}}T}}) \, . \tag{3}$$

Let’s present the frequency dependence of the spectral emissivity of a real-body as a polynomial

$$\varepsilon(\nu,T)=\sum_{i=-3}^{\infty}a_i(T)\nu^i \quad , \tag{4}$$

where $a_i$ are the coefficients.

The total energy density of a real-body radiation in the finite frequency range of the spectrum is defined as:

$$I(\nu_1,\nu_2,T)=\frac{8\pi h}{c^3}\int_{\nu_1}^{\nu_2}\nu^3\varepsilon(\nu,T)\mathrm{Li}_0(e^{-\frac{h\nu}{k_\mathrm{B}T}})d\nu\,. \tag{5}$$

Using the relationship between the total energy density Eq. (5) and the total radiation power per unit area $I^{\mathrm{SB}}=\frac{c}{4}I$, the Stefan-Boltzmann law in the finite frequency range of the spectrum takes the form

$$I^{\mathrm{SB}}(\nu_1,\nu_2,T)=\frac{2\pi h}{c^2}\int_{\nu_1}^{\nu_2}\nu^3\varepsilon(\nu,T)\mathrm{Li}_0(e^{-\frac{h\nu}{k_\mathrm{B}T}})d\nu\,. \tag{6}$$

The total emissivity is represented as:

$$\varepsilon(\nu_1,\nu_2,T)=\frac{I(\nu_1,\nu_2,T)}{I^{\mathrm{BB}}(\nu_1,\nu_2,T)} \quad , \tag{7}$$

where

$$I^{\mathrm{BB}}(\nu_1,\nu_2,T)=\frac{48\pi(k_\mathrm{B}T)^4}{c^3h^3}[P_3(x_1)-P_3(x_2)] \tag{8}$$

is the total energy density of blackbody radiation in the finite frequency range of the spectrum [48]. Here $x=\frac{h\nu}{k_\mathrm{B}T}$ and $P_3(x)$ is defined as:

$$P_3(x)=\sum_{s=0}^{3}\frac{(x)^s}{s!}\mathrm{Li}_{4-s}(e^{-x})\,, \tag{9}$$

where

$$\mathrm{Li}_{4-s}(e^{-x})=\sum_{k=1}^{\infty}\frac{e^{-kx}}{k^{4-s}} \quad , \left|e^{-kx}\right|<1 \tag{10}$$

is the polylogarithm function of the order 4-*s* [47].

According to [46], the number density of photons of a real-body radiation with a photon energy from $h\nu_1$ to $h\nu_2$ is represented in the form

$$n=\frac{8\pi}{c^3}\int_{\nu_1}^{\nu_2}\varepsilon(\nu,T)\nu^2\mathrm{Li}_0(e^{-\frac{h\nu}{k_\mathrm{B}T}})d\nu\ . \tag{11}$$

The free energy density in the finite frequency range of the spectrum is defined as [46]:

$$a(\nu_1,\nu_2,T)=\frac{8\pi k_\mathrm{B}}{c^3}\int_{\nu_1}^{\nu_2}\nu^2\varepsilon(\nu,T)\ln((1-e^{-\frac{h\nu}{k_\mathrm{B}T}})d\nu\ . \tag{12}$$

The thermodynamic functions of a real-body radiation in a finite range of frequencies are defined by the following expressions [46]:

1) Entropy density $s=\frac{S}{V}$:

$$s=-\frac{\partial a}{\partial T}; \tag{13}$$

2) Heat capacity at constant volume per unit volume $c_V=\frac{C_V}{V}$:

$$c_V=\left(\frac{\partial I(\nu_1,\nu_2,T)}{\partial T}\right)_V; \tag{14}$$

3) Pressure of photons per volume $p=\frac{P}{V}$:

$$p=-a\ . \tag{15}$$

**3. Polylogarithmic representation of thermal radiative properties of a real-body**

To compute the total energy density for a given temperature over the finite frequency range of the spectrum, it is necessary to compute the integral in Eq. (5). The integral can be integrated by parts to give

$$I(\nu_1,\nu_2,T)=\frac{8\pi(k_BT)^4}{c^3h^3}\sum_{i=-3}^{\infty}a_i(3+i)!\left(\frac{k_\mathrm{B}T}{h}\right)^i A_i(x_1,x_2)\ , \tag{16}$$

where

$$A_i(x_1,x_2)=P_{3+i}(x_1)-P_{3+i}(x_2)\ . \tag{17}$$

Here $x=\frac{h\nu}{k_\mathrm{B}T}$ and $P_3(x)$ is defined as:

$$P_{3+i}(x)=\sum_{s=0}^{3+i}\frac{(x)^s}{s!}\mathrm{Li}_{4+i-s}(e^{-x}), \qquad (18)$$

where

$$\mathrm{Li}_{4+i-s}(e^{-x})=\sum_{k=1}^{\infty}\frac{e^{-kx}}{k^{4+i-s}} \quad , \left|e^{-kx}\right|<1 \qquad (19)$$

is the polylogarithm function of the order $4+i$-$s$ [47].

In accordance with Eq. (6), the Stefan-Boltzmann law in the finite frequency range of the spectrum takes the form

$$I^{\mathrm{SB}}(v_1,v_2,T)=\frac{2\pi(k_B T)^4}{c^2h^3}\sum_{i=-3}^{\infty}a_i(3+i)!\left(\frac{k_{\mathrm{B}}T}{h}\right)^i\left[P_{3+i}(x_1)-P_{3+i}(x_2)\right]. \qquad (20)$$

The total radiation power $I_{\mathrm{total}}$ emitted by a heated surface area $S$ of a sample is defined as:

$$I_{\mathrm{total}}=SI^{\mathrm{SB}}(v_1,v_2,T). \qquad (21)$$

In accordance with Eq. (7), Eq. (8), Eq. (16) and Eq. (17), the total emissivity can be presented in the form

$$\varepsilon(v_1,v_2,T)=\frac{\sum_{i=-3}^{\infty}a_i(T)(3+i)!\left(\frac{k_{\mathrm{B}}T}{h}\right)^i\left[P_{3+i}(x_1)-P_{3+i}(x_2)\right]}{6\left[P_3(x_1)-P_3(x_2)\right]}. \qquad (22)$$

Using Eq. (4) and after computing the integral in Eq. (11), the polylogarithmic representation of the number density of photons can be written as:

$$n=\frac{8\pi}{c^3}\sum_{i=-2}^{\infty}a_i\left(\frac{k_{\mathrm{b}}T}{h}\right)^{3+i}(2+i)!\,B_i(x_1,x_2), \qquad (23)$$

where

$$B_i(x_1,x_2)=P_{2+i}(x_1)-P_{2+i}(x_2). \qquad (24)$$

In conclusion of this paragraph, it is essential to note that the analytical expressions derived above in the case of black-body radiation, when $a_i=0$ and $a_0=1$, take a well-known expressions [48].

## 4. Thermodynamics of a real-body radiation

After computing the integral in Eq. (12), the general expressions for the thermodynamic functions of a real-body radiation Eqs. (12) - (15) can be expressed in terms of the polylogarithm functions as follows:

(1) Helmholtz free energy density $a$ :

$$a = -\frac{8\pi k_B{}^4}{c^3 h^3} T^4 \sum_{i=-2}^{\infty} a_i (2+i)! \left(\frac{k_\mathrm{B} T}{h}\right)^i C_i(x_1, x_2), \tag{25}$$

where

$$C_i(x_1, x_2) = \left\{ \left[P_{3+i}(x_1) - P_{3+i}(x_2)\right] - \frac{1}{(3+i)!}\left(x_1^{3+i}\mathrm{Li}_1(e^{-x_1}) - x_2^{3+i}\mathrm{Li}_1(e^{-x_2})\right)\right\}. \tag{26}$$

(2) Entropy density $s$ :

$$s = \frac{8\pi k_B{}^4}{c^3 h^3} T^3 \sum_{i=-2}^{\infty} a_i \left(\frac{k_\mathrm{B} T}{h}\right)^i (2+i)!(4+i) D_i(x_1, x_2), \tag{27}$$

where

$$D_i(x_1, x_2) = \left\{ [P_{3+i}(x_1) - P_{3+i}(x_2)] - \frac{1}{(4+i)!}[x_1^{3+i}\mathrm{Li}_1(e^{-x_1}) - x_2^{3+i}\mathrm{Li}_1(e^{-x_2})]\right\}. \tag{28}$$

(3) Heat capacity at constant volume per volume, $c_V$ :

$$c_V = \frac{8\pi k_B{}^4}{c^3 h^3} T^3 \sum_{i=-2}^{\infty} a_i \left(\frac{k_\mathrm{B}}{h}\right)^i T^i (4+i)! E_i(x_1, x_2), \tag{29}$$

where

$$E_i(x_1, x_2) = \left\{ [P_{3+i}(x_1) - P_{3+i}(x_2)] + \frac{1}{(4+i)!}[x_1^{4+i}\mathrm{Li}_0(e^{-x_1}) - x_2^{4+i}\mathrm{Li}_0(e^{-x_2})]\right\}. \tag{30}$$

(4) Pressure $p$:

$$p = \frac{8\pi k_B{}^4}{c^3 h^3} T^4 \sum_{i=-2}^{\infty} a_i (2+i)! \left(\frac{k_\mathrm{B} T}{h}\right)^i C_i(x_1, x_2) \ , \tag{31}$$

where

$$C_i(x_1,x_2)=\left\{\left[P_{3+i}(x_1)-P_{3+i}(x_2)\right]-\frac{1}{(3+i)!}\left(x_1^{3+i}\mathrm{Li}_1(e^{-x_1})-x_2^{3+i}\mathrm{Li}_1(e^{-x_2})\right)\right\} \quad . \tag{32}$$

By definition [46], $a=u-Ts$, (where $u$ is the internal energy density), we obtain the analytical expression for $u$

$$u(x_1,x_2,T)=a+Ts=\frac{8\pi(k_BT)^4}{c^3h^3}\sum_{i=-3}^{\infty}a_i(3+i)!\left(\frac{k_BT}{h}\right)^i A_i(x_1,x_2) \quad , \tag{33}$$

where $A(x_1,x_2)$ is defined by Eq. (17).

The enthalpy density $h$ follows from its definition, $h=u+p$, giving

$$h(x_1,x_2,T)=\frac{8\pi k_B{}^4}{c^3h^3}T^4\sum_{i=-2}^{\infty}a_i\left(\frac{k_BT}{h}\right)^i(2+i)!(4+i)D_i(x_1,x_2) \quad . \tag{34}$$

The Gibbs free energy density $g$, by definition, is $h-Ts$, thus

$$g(x_1,x_2,T)=0 \ . \tag{35}$$

The chemical potential density $\mu=\left(\frac{\partial g}{\partial n}\right)_{T,V}$, as seen from Eq. (35), is zero

$$\mu(x_1,x_2,T)=0 \ . \tag{36}$$

In conclusion, it should be noted that the obtained analytical expressions for the thermal radiative and thermodynamic functions of a real-body radiation in a finite range of frequencies can easily be presented in the wavenumber ($\tilde{\nu}$) domain. In this case, we should use the following relationships [49]:

$$\nu=c\tilde{\nu} \tag{37}$$

$$d\nu=cd\tilde{\nu} \tag{38}$$

$$\int_{\nu_1}^{\nu_2}I^{\mathrm{P}}(\nu,T)d\nu=\int_{\tilde{\nu}_1}^{\tilde{\nu}_2}I^{\mathrm{P}}(\tilde{\nu},T)d\tilde{\nu} \ . \tag{39}$$

Note that using different spectral units produces the same result, because it represents the same physical quantity.

## 5. Thermal radiative and thermodynamic properties of liquid and solid zirconium carbide

Now let us consider an example related to the study of thermal radiative and thermodynamic properties of liquid and solid zirconium carbide using experimental data for the normal spectral emissivity in the visible-near infrared range at melting/freezing temperature.

It is well-known that the rapid development of space and missile technologies requires ultra-high temperature ceramics with the melting temperature up to 4273 K [50, 51]. Zirconium carbide is a good candidate material for using it in environments with extreme temperatures. Some application of ZrC are: a) nuclear fuel coating in high temperature Generation IV reactors [52]; b) thermal shield in aerospace applications [53]; c) solar energy receiver with low emissivity and high absorptivity [54]; etc. Thus, the investigation of the thermal radiative and thermodynamic properties of zirconium carbide under extreme conditions is a research domain of great interest both for basic science and industrial applications.

In [41], the radiance spectra of zirconium carbide were measured in the frequency range ($0.333\,\mathrm{PHz} \le \nu \le 0.545\,\mathrm{PHz}$) at temperature $T$ = 3155 K during the melting and freezing arrests while cooling and heating the samples. The measured normal spectral emissivity of solid and liquid zirconium carbide is approximated by the following analytical expression:

$$\varepsilon(\nu,T) = \sum_{i=-2}^{0} \tilde{a}_i \nu^i = \tilde{a}_0 + \tilde{a}_{-1}\nu^{-1} + \tilde{a}_{-2}\nu^{-2} \quad , \tag{40}$$

where

$$\text{Solid ZrC: } \tilde{a}_0 = 0.6968 \; ; \tilde{a}_{-1} = -8.2503 \times 10^{7}\,\mathrm{Hz}; \tilde{a}_{-2} = 1.4739 \times 10^{16}\,\mathrm{Hz}^2 \tag{41}$$

$$\text{Liquid ZrC: } \tilde{a}_0 = 0.75746 \; ; \tilde{a}_{-1} = -1.40363 \times 10^{14}\,\mathrm{Hz}; a_{-2} = 1.66387 \times 10^{29}\,\mathrm{Hz}^2 . \tag{42}$$

The gap between the normal spectral emissivity of solid and liquid ZrC is observed in the spectral frequency range from $0.333\,\mathrm{PHz}$ to $0.545\,\mathrm{PHz}$ [41]. The normal spectral emissivity of both solid and liquid zirconium carbide increase with increasing $\nu$.

Using the general expressions for the thermal radiative and thermodynamic functions of a real-body obtained above and Eq. (40), in the case of zirconium carbide, we obtain:

1. The total energy density in the finite frequency range of the spectrum:

$$I = \tilde{a}_0 I_0 + \tilde{a}_{-1} I_{-1} + \tilde{a}_{-2} I_{-2}, \tag{43}$$

where

$$I_0 = \frac{48\pi(k_B T)^4}{c^3 h^3}\left[P_3(x_1) - P_3(x_2)\right], \tag{44}$$

$$I_{-1} = \frac{16\pi(k_B T)^3}{c^3 h^2}\left[P_2(x_1) - P_2(x_2)\right], \tag{45}$$

$$I_{-2} = \frac{8\pi(k_B T)^2}{c^3 h}\left[P_1(x_1) - P_1(x_2)\right]. \tag{46}$$

2. The total radiation power per unit area in the finite frequency range (Stefan-Boltzmann law):

$$I^{\mathrm{SB}} = \tilde{a}_0 I^{\mathrm{SB}}{}_0 + \tilde{a}_{-1} I^{\mathrm{SB}}{}_{-1} + \tilde{a}_{-2} I^{\mathrm{SB}}{}_{-2}, \tag{47}$$

where

$$I^{\mathrm{SB}}{}_0 = \frac{12\pi(k_B T)^4}{c^2 h^3}\left[P_3(x_1) - P_3(x_2)\right], \tag{48}$$

$$I^{\mathrm{SB}}{}_{-1} = \frac{4\pi(k_B T)^3}{c^2 h^2}\left[P_2(x_1) - P_2(x_2)\right], \tag{49}$$

$$I^{\mathrm{SB}}{}_{-2} = \frac{2\pi(k_B T)^2}{c^2 h}\left[P_1(x_1) - P_1(x_2)\right]. \tag{50}$$

3. Total emissivity:

$$\varepsilon(v_1, v_2, T) = \frac{\tilde{a}_0 I_0 + \tilde{a}_{-1} I_{-1} + \tilde{a}_{-2} I_{-2}}{(3!)[P_3(x_1) - P_3(x_2)]}.$$

4. Number density of photons with a photon energy from $hv_1$ to $hv_2$:

$$n = \tilde{a}_0 n_0 + \tilde{a}_{-1} n_{-1} + \tilde{a}_{-2} n_{-2}. \tag{51}$$

where

$$n_0 = \frac{16\pi k_B{}^3}{c^3 h^3} T^3 \{[P_2(x_1) - P_2(x_2)]\}, \tag{52}$$

$$n_{-1} = \frac{8\pi k_B{}^2}{c^3 h^2} T^2 \{[P_1(x_1) - P_1(x_2)]\}, \tag{53}$$

$$n_{-2} = \frac{8\pi k_B}{c^3 h} T \{[P_0(x_1) - P_0(x_2)]\}. \tag{54}$$

5. Helmholtz free energy density $a$ :

$$a = \tilde{a}_0 a_0 + \tilde{a}_{-1} a_{-1} + \tilde{a}_{-2} a_{-2}, \tag{55}$$

where

$$a_0 = -\frac{16\pi k_B{}^4}{c^3 h^3} T^4 \left\{ [P_3(x_1) - P_3(x_2)] - \frac{1}{6}\left( x_1^3 \mathrm{Li}_1(e^{-x_1}) - x_2^3 \mathrm{Li}_1(e^{-x_2}) \right) \right\}, \tag{56}$$

$$a_{-1} = -\frac{8\pi k_B{}^3}{c^3 h^2} T^3 \left\{ [P_2(x_1) - P_2(x_2)] - \frac{1}{2}\left( x_1^2 \mathrm{Li}_1(e^{-x_1}) - x_2^2 \mathrm{Li}_1(e^{-x_2}) \right) \right\}, \tag{57}$$

$$a_{-2} = -\frac{8\pi k_B{}^2}{c^3 h} T^2 \left\{ [P_1(x_1) - P_1(x_2)] - \left( x_1 \mathrm{Li}_1(e^{-x_1}) - x_2 \mathrm{Li}_1(e^{-x_2}) \right) \right\}. \tag{58}$$

6. Entropy density $s$:

$$s = \tilde{a}_0 s_0 + \tilde{a}_{-1} s_{-1} + \tilde{a}_{-2} s_{-2}, \tag{59}$$

where

$$s_0 = \frac{64\pi k_B{}^4}{c^3 h^3} T^3 \left\{ [P_3(x_1) - P_3(x_2)] - \frac{1}{24}\left( x_1^3 \mathrm{Li}_1(e^{-x_1}) - x_2^3 \mathrm{Li}_1(e^{-x_2}) \right) \right\}, \tag{60}$$

$$s_{-1} = \frac{24\pi k_B{}^3}{c^3 h^2} T^2 \left\{ [P_2(x_1) - P_2(x_2)] - \frac{1}{6}\left( x_1^2 \mathrm{Li}_1(e^{-x_1}) - x_2^2 \mathrm{Li}_1(e^{-x_2}) \right) \right\}, \tag{61}$$

$$s_{-2} = \frac{16\pi k_B{}^2}{c^3 h} T \left\{ [P_1(x_1) - P_1(x_2)] - \frac{1}{2}\left( x_1 \mathrm{Li}_1(e^{-x_1}) - x_2 \mathrm{Li}_1(e^{-x_2}) \right) \right\}. \tag{62}$$

7. Heat capacity at constant volume per unit volume $c_V$ :

$$c_V = \tilde{a}_0 c_{V0} + \tilde{a}_{-1} c_{V-1} + \tilde{a}_{-2} c_{V-2}\,, \tag{63}$$

where

$$c_{V0} = \frac{192\pi k_B^{\;4}}{c^3 h^3} T^3 \left\{ \left[ P_3(x_1) - P_3(x_2) \right] + \frac{1}{24} \left( x_1^4 \mathrm{Li}_0(e^{-x_1}) - x_2^4 \mathrm{Li}_0(e^{-x_2}) \right) \right\}, \tag{64}$$

$$c_{V-1} = \frac{48\pi k_B^{\;3}}{c^3 h^2} T^2 \left\{ \left[ P_2(x_1) - P_2(x_2) \right] + \frac{1}{6} \left( x_1^3 \mathrm{Li}_0(e^{-x_1}) - x_2^3 \mathrm{Li}_0(e^{-x_2}) \right) \right\}, \tag{65}$$

$$c_{V-2} = \frac{16\pi k_B^{\;2}}{c^3 h^1} T \left\{ \left[ P_1(x_1) - P_1(x_2) \right] + \frac{1}{2} \left( x_1^2 \mathrm{Li}_0(e^{-x_1}) - x_2^2 \mathrm{Li}_0(e^{-x_2}) \right) \right\}. \tag{66}$$

The calculated values of thermal radiative and thermodynamic functions of thermal radiation of solid and liquid ZrC emitted by a heated surface per unit area of the sample are presented in Table 1.

Now let us calculate thermal radiative and thermodynamic properties of thermal radiation of solid and liquid zirconium carbide emitted by a surface area *S* of the sample.

According to [41], the zirconium carbide sample under investigation is a disc about 1 mm thick and around 10 mm in diameter. Then, in accordance with Eq. (21), the total radiation power emitted by a surface area $S$ of the liquid and solid ZrC sample are defined as

$$\text{Solid ZrC:} \;\; I^{\mathrm{SB}}_{\mathrm{Solidtotal}}(T) = S\, I^{\mathrm{SB}}_{\mathrm{Solid}}(\nu_1, \nu_2, T) \tag{67}$$

$$\text{Liquid ZrC:} \;\; I^{\mathrm{SB}}_{\mathrm{Liquidtotal}}(T) = S\, I^{\mathrm{SB}}_{\mathrm{Liquid}}(\nu_1, \nu_2, T)\;, \tag{68}$$

where $I^{\mathrm{SB}}_{\mathrm{Liquid}}(\nu_1, \nu_2, T)$ and $I^{\mathrm{SB}}_{\mathrm{Solid}}(\nu_1, \nu_2, T)$ are the total radiation power emitted by a heated surface per unit area of the zirconium carbide sample in the finite frequency range. Their values are presented in Table 1.

The surface area *S* of the zirconium carbide sample can be written as:

$$S = 2\pi \left( \frac{d}{2} \right)^2 + \pi h d = 1.885 \times 10^{-4}\,\mathrm{m}^2\,. \tag{69}$$

Then, in accordance with Eq. (32) and Table 1, we have

$$\text{Solid ZrC:} \quad I^{\mathrm{SB}}_{\mathrm{Solidtotal}} = S\, I^{\mathrm{SB}}(\nu_1, \nu_2, T) = 7.379 \times 10^{2}\ \mathrm{W} \tag{70}$$

$$\text{Liquid ZrC:} \quad I^{\mathrm{SB}}_{\mathrm{Liquidtotal}} = S\, I^{\mathrm{SB}}(\nu_1, \nu_2, T) = 9.178 \times 10^{2}\ \mathrm{W}\ . \tag{71}$$

A volume of the ZrC sample under study can be calculated by the following expression:

$$V = \pi\, h \left(\frac{d}{2}\right)^{2} = 7.854 \times 10^{-8}\,\mathrm{m}^{3} \quad . \tag{72}$$

Then, in accordance with Eq. (18), Eq. (19), and Table 1, the total energy densities of liquid and solid zirconium carbide sample in the finite frequency range have the following values:

$$\text{Solid ZrC:} \quad I_{\mathrm{Solidtotal}}\,(T) = V\, I(\nu_1, \nu_2, T) = 4.102 \times 10^{-9}\,\mathrm{J} \tag{73}$$

$$\text{Liquid ZrC:} \quad I_{\mathrm{Liquidtotal}}\,(T) = V\, I(\nu_1, \nu_2, T) = 5.102 \times 10^{-9}\,\mathrm{J}\ . \tag{74}$$

The total numbers of photons $N$ emitted by liquid and solid ZrC in the finite frequency range $0.15\,\mathrm{THz} \le \nu \le 2.88\,\mathrm{THz}$ at temperature $T$ = 3155 K have the following values:

$$\text{Solid ZrC:} \quad N_{\mathrm{Solid}} = Vn = 3.486 \times 10^{10} \tag{75}$$

$$\text{Liquid ZrC:} \quad N_{\mathrm{Liquid}} = Vn = 4.336 \times 10^{10}\ . \tag{76}$$

Now let us calculate the thermodynamic functions of thermal radiation of liquid and solid ZrC emitted by a heated surface area $S$ of the sample. Using Eq. (39) and Table 1, for the total free energy $A_{\mathrm{total}}$ , we obtain

$$\text{Solid ZrC:} \quad A_{\mathrm{Solidtotal}} = a\, V = -1.367 \times 10^{-9}\,\mathrm{J} \tag{77}$$

$$\text{Liquid ZrC:} \quad A_{\mathrm{Liquidtotal}} = a\, V = -1.701 \times 10^{-9}\,\mathrm{J}\ . \tag{78}$$

In Table 2, the calculated values of thermal radiative and thermodynamic functions of thermal radiation of liquid and solid zirconium carbide emitted by a heated surface area $S$ of the sample are presented in the finite frequency range from $0.333\,\mathrm{PHz}$ to $0.545\,\mathrm{PHz}$ $0.550\,\mu\mathrm{m}$ at the eutectic melting (freezing) temperature $T$ = 3155 K. As it can be clearly seen, the gaps between the thermal radiative and thermodynamic functions of liquid and solid zirconium carbide are observed. The existence of these gaps indicates that liquid ZrC has probably a more metallic nature than solid ZrC [41].

## 6. Conclusions

In this work, the general analytical expressions for the thermal radiative and thermodynamic functions of a real-body are obtained in the finite frequency range using the frequency dependence of the normal spectral emissivity in the form of $\varepsilon(\nu,T)=\sum_{i=-3}^{m} a_i \nu^i$ . In this case, the Stefan-Boltzmann law, total energy density, number density of photons, Helmholtz free energy density, enthalpy density, internal energy density, entropy density, heat capacity at constant volume, and pressure are expressed in the terms of the polylogarithm functions. This polylogarithmic representation allows to calculate the thermal radiative and thermodynamic functions analytically.

The analytical expressions obtained in this work have been applied to the study of thermal radiative and thermodynamic properties of solid and liquid zirconium carbide using the experimental data for the frequency dependence of the normal spectral emissivity at melting (freezing) point. The calculated values of the total radiation power per unit area, total energy density, number density of photons, Helmholtz free energy density, enthalpy density, internal energy density, entropy density, heat capacity at constant volume, and pressure in a spectral range $0.550\,\mu\text{m} \le \lambda \le 0.900\,\mu\text{m}$ at temperature $T$ = 3155 K are presented in Table 1.

In Table 2, the thermal radiative and thermodynamic functions of thermal radiation of solid and liquid ZrC emitted by a heated surface area $S$ of the sample are presented. The existence of the gaps between the thermal radiative and thermodynamic functions of solid ZrC and that of liquid ZrC in the visible range are confirmed.

In conclusion, it is important to note that there are several classes of materials and space objects for which the thermal radiative and thermodynamic properties can be described using polylogarithm functions. These real-bodies are: a) luminous flames [31]; b) cobalt, iron, and nickel at the melting points [42, 43]; c) the Milky Way and other galaxies [44, 45, 55]; etc. As a result, analytical expressions for the thermal radiative and thermodynamic functions of a real-body radiation in various frequency ranges at different temperatures can be obtained.

These and other topics will be points of discussion in subsequent publications.

**Acknowledgments**

The authors cordially thank Professor N.P. Malomuzh for fruitful discussions.

## References

1. Michalski L, Eckersdorf K, McGhee J. Temperature measurements. Chichester: Wiley; 1991.
2. Magison EC. Temperature measurements in industry. Research Triangle Park: ISA; 1990.
3. Fukuyama H, (Editor), Waseda Y. (Editor). High-temperature measurements of materials (Advances in Materials Research). Springer; 2008.
4. Riethof T, Acchione B, Branyan E. High-temperature spectral emissivity studies on some refractory metals and carbides. In: Temperature, its measurement and control in science and industry. New York: Reinhold Publishing Corporation; 1962.
5. Meng S, Chen H, Hu J, Wang Z. Radiative properties characterization of ZrB2-SiC-based ultrahigh temperature ceramic at high temperature. Materials and Design 2011; 32: 377-81.
6. Bober M, Karow HU, Muller K. Study of the spectral reflectivity and emissivity of liquid ceramics. High Temperatures – High Pressures 1980; 12: 161-68.
7. Soerensen DD, Clausen S, Mercer JB, Pedersen LJ. Determining the emissivity of pig skin for accurate infrared thermography. Computers and Electronics in Agriculture 2014; 109: 52–58.
8. Ke R, Zhang Y, Zhou Y. Study on infrared emissivity of biomimetic motheye sapphire single crystal. Optik - International Journal for Light and Electron Optics 2014; 125: 6991-994.
9. Adibekyan A, Monte C, Kehrt M, Gutschwager B, Hollandt J. Emissivity measurement under vacuum from to and from to at PTB. International Journal of Thermophysics 2015; 36: 283-89.
10. Tairan F, Peng T, Minghao D. Simultaneous measurements of high-temperature total hemispherical emissivity and thermal conductivity using a steady-state calorimetric technique. Measurement Science and Technology. Article id. 015003 (2015).
11. Nakazawa K, Ohnishi A. Simultaneous measurement method of normal spectral emissivity and optical constants of solids at high temperature in vacuum. International Journal of Thermophysics 2010; 31: 2010-018.
12. Barducci A, Guzzi D, Marcoionni P, Pippi I. A new algorithm for temperature and spectral emissivity retrieval over active fires in the TIR spectral range. IEEE Transactions on Geoscience and Remote Sensing 2004; 42: 1521-29.
13. Matsumoto T, Cezairliyan A, Basak D. Hemispherical total emissivity of niobium, molybdenum, and tungsten at high temperatures using a combined transient and brief steady-state technique. International Journal of Thermophysics 1999; 20: 943–52.

14. Cagran C, Pottlacher G. Thermophysical properties and normal spectral emittance of Iridium up to 3500 K. International Journal of Thermophysics 2007; 28: 697-710.
15. Matsumoto T, Cezairliyan A, Basak D. Hemispherical total emissivity of niobium, molybdenum, and tungsten at high temperatures using a combined transient and brief steady-state technique. International Journal of Thermophysics 1999; 20: 943–52.
16. Krishnan S, Hansen GP, Hauge RH, Margrave JL. Spectral emissivities and optical properties of electromagnetically levitated liquid metals as functions of temperature and wavelength. High Temperature Science 1990; 29: 17– 52.
17. Krishnan S, Nordine PC. Spectral emissivities in the visible and infrared of liquid Zr, Ni, and nickel-based binary alloys Journal of Applied Physics. 1996; 80: 1735-42.
18. Touloukian YS, DeWitt DP. Thermal radiative properties: Metallic elements and alloys, Thermophysical Properties of Matter, Vol. 7. New York, Washington: IFI/Plenum; 1970.
19. Watmough DJ, Oliver R. Emissivity of human skin in the waveband between 2µ and 6µ. Nature 1968; 219: 622-24.
20. Wen C-D, Mudawar I. Emissivity characteristics of roughened aluminum alloy surfaces and assessment of multispectral radiation thermometry (MRT) emissivity models. International Journal of Heat and Mass Transfer 2004; 47: 3591–605.
21. Tsai BK, Shemaker RL, DeWitt DP, Cowans BA, Dardas Z, Delgass WN, Dail GJ: Dual-wavelength radiation thermometry: Emissivity compensation algorithms International Journal of Thermophysics 1990; 11: 269-81.
22. Hopkins MF. Four color pyrometer for metal emissivity characterization. SPIE—The International Society of Optical Engineering 1996; 2599: 294-301.
23. Khan MA, Allemand CD, Eager TW. Noncontact temperature measurement. II. Interpolation based techniques. Review Scientific Instrument 1991; 62: 403-09.
24. Duvaut Th, Georgeault D, Beaudoin JL. Multiwavelength infrared pyrometry: Optimization and computer simulations, Infrared Physics & Technology 1995; 36: 1089-109.
25. Gardner JL, Jones TP, Davies MR. A six-wavelength radiation pyrometer," High Temperatures—High Pressures 1981; 13: 459-66.
26. Gardner JL. Computer modeling of multiwavelength pyrometer for measuring true surface temperature. High Temperatures—High Pressures 1980; 12: 699-705.
27. Hoch M. The integral six-color pyrometer: A new general method of temperature determination.

High Temperatures—High Pressures 1992; 24: 607-23.

28. Hoch M. The integral six-color pyrometer: Linear dependence of the radiance temperature Tr on the wavelength lambda. Review of Scientific Instruments 1992; 63: 2274-281.
29. Gathers GR. Monte Carlo studies of multiwavelength pyrometry using linearized equations. International Journal of Thermophysics 1992; 13: 361-82.
30. Gathers GR. Analysis of multiwavelength pyrometry using nonlinear chi-square fits and Monte Carlo methods. International Journal of Thermophysics 1992; 13: 539-54.
31. Siegel R, Howell JR. Thermal radiation heat transfer. New York: McGraw-Hill; 1972
32. Edwards DK, Catton I. Advances in the thermophysical properties at extreme temperatures and pressures. In: Proceedings of third symposium on thermophysical properties 1965: 189-99
33. Fisenko AI, Ivashov SN. Determination of the true temperature of emitted radiation bodies from generalized Wien's displacement law. Journal Physics D: Applied Physics 1999; 32: 2882-85.
34. Fisenko AI, Ivashov SN. Optical and radiant characteristics of tungsten at high temperatures. Journal of Engineering Physics 1989; 57: 838-41.
35. Ershov VA, Fisenko AI. Radiative characteristics of flames. Combustion, Explosion and Shock Waves 1992; 28: 159-61.
36. Fisenko AI, Ivashov SN. Determination of the true temperature of molybdenum and luminous flames from generalized Wien's displacement and Stefan–Boltzmann's jaws: Thermodynamics of thermal radiation. International Journal of Thermophysics 2009; 30: 1524-35.
37. Fisenko AI, Lemberg V. Radiative properties of stoichiometric hafnium, titanium, and zirconium carbides: Thermodynamics of thermal radiation. International Journal of Thermophysics 2012; 33: 513-27.
38. Fisenko AI, Lemberg V. Generalized Wien's displacement law in determining the true temperature of $ZrB_2$–SiC-based ultrahigh-temperature ceramic: Thermodynamics of thermal radiation. International Journal of Thermophysics 2013; 34: 486-95.
39. Fisenko AI, Lemberg V. Polylogarithmic representation of radiative and thermodynamic properties of thermal radiation in a given spectral range: I. Black-body radiation. International Journal of Thermophysics 2013. In review.
40. Manara D, Bruycker F. De, Boboridis K, Tougait O, Eloirdy R, Malki M. High temperature radiance spectroscopy measurements of solid and liquid uranium and plutonium carbides. Journal of Nuclear Materials 2012; 426: 126-38.

41. Manara D, Jackson HF, Perinetti-Casoni C, Boboridis K, Welland MJ, Luzzi L, Ossi PM, Lee WE. The ZrC-C eutectic structure and melting behavior: A high-temperature radiance spectroscopy study. Journal of the European Ceramic Society 2013; 33: 1349-61.
42. Watanabe H, Susa M, Fukuyama H, Nagata K. Phase dependence (liquid/solid) of normal spectral emissivities of noble metals at melting points. International Journal of Thermophysics 2003; 24: 223-37.
43. Watanabe H, Susa M, Fukuyama H, Nagata K. Phase (liquid/solid) dependence of the normal spectral emissivity for iron, cobalt, and nickel at melting points. International Journal of Thermophysics 2003; 24: 473-88.
44. Reach WT, Dwek E, Fixsen D J, Hewagama T, Mather J C, Shafer R A, Banday A J, Bennett C L, Cheng ES, Eplee Jr R E, Leisawitz D, Lubin P M, Read SM, Rosen LP, Shuman FGD, Smoot GF, Sodroski T J, Wright EL. Far-Infrared Spectral Observations of the Galaxy by COBE. Astrophysical Journal 1995; 451: 188-99.
45. Spinoglio L, Malkan, MA, Smith HA, González-Alfonso E, Fischer J. The far-infrared emission line and continuum spectrum of the Seyfert galaxy NGC 1068. Astrophysical Journal 2005; 623: 123-36.
46. Landau LD, Lifshitz EM. Statistical physics, course of theoretical physics, vol. 5. Oxford, New York: Pergamon Press; 1980.
47. Abramowitz M, Stegun IA. Handbook of mathematical functions with formulas, graphs, and mathematical tables. New York: Dover Publications; 1972.
48. Fisenko AI, Lemberg V. On the radiative and thermodynamic properties of the cosmic radiations using COBE FIRAS instrument data: I. Cosmic microwave background radiation. Astrophysics and Space Science 2014; 352: 221-30.
49. http://en.wikipedia.org/wiki/Wavenumber
50. Wuchina E, Opila E, Opeka M, Fahrenholtz W, Talmy I. UHTCs: Ultra-high temperature ceramic materials for extreme environment applications. Interface 2007; 16: 30-6.
51. Shaffer PTB. Engineering properties of carbides. In Engineered materials handbook, Vol. 4, Ceramics and Glass. Metals Park, OH: ASM International; 1991.
52. Doonyapong W. Performance modeling of Deep Burn TRISO fuel using ZrC as a load-bearing layer and an oxygen getter. Journal of Nuclear Materials 2010; 396: 149–58.
53. Levine SR, Opila EJ, Halbig MC, Kiser JD, Singh M, Salem JA. Evaluation of ultra-high

temperature ceramics for aeropropulsion use. Journal of the European Ceramic Society 2002; 22: 2757–67.

54. Sani E, Mercatelli L, Francini F, Sans JL, Sciti D. Ultra-refractory ceramics for high-temperature solar absorbers. Scripta Mater 2011; 65: 75–8.

55. Fisenko AI, Lemberg V. On the radiative and thermodynamic properties of the cosmic radiations using *COBE* FIRAS instrument data: III. Galactic far-infrared radiation. Research in Astronomy and Astrophysics 2015; In press.

| **Quantity** | **Zirconium Carbide:** *Solid Phase* | **Zirconium Carbide:** *Liquid Phase* | **Gaps** Between *Liquid and Solid Phases* |
|---|---|---|---|
| $I(\nu_1,\nu_2,T)$ $\left[\mathrm{J\ m^{-3}}\right]$ | $5.223\times10^{-2}$ | $6.496\times10^{-2}$ | $1.273\times10^{-2}$ |
| $I^{\mathrm{SB}}\ (\nu_1,\nu_2,T)$ $\left[\mathrm{W\ m^{-2}}\right]$ | $3.915\times10^{6}$ | $4.869\times10^{6}$ | $0.954\times10^{6}$ |
| $\varepsilon$ | 0.697 | 0.867 | 0.170 |
| $a$ $\left[\mathrm{J\ m^{-3}}\right]$ | $-1.741\times10^{-2}$ | $-2.165\times10^{-2}$ | $-0.424\times10^{-2}$ |
| $s$ $\left[\mathrm{J\ m^{-3}\ K^{-1}}\right]$ | $2.207\times10^{-5}$ | $2.745\times10^{-5}$ | $0.538\times10^{-5}$ |
| $p$ $\left[\mathrm{J\ m^{-3}}\right]$ | $1.741\times10^{-2}$ | $2.165\times10^{-2}$ | $0.424\times10^{-2}$ |
| $c_V$ $\left[\mathrm{J\ m^{-3}\ K^{-1}}\right]$ | $6.622\times10^{-5}$ | $8.236\times10^{-5}$ | $1.614\times10^{-5}$ |
| $n$ $[\mathrm{m^{-3}}]$ | $4.439\times10^{17}$ | $5.521\times10^{17}$ | $1.082\times10^{17}$ |

**Table 1** Calculated values of the thermal radiative and thermodynamic functions of thermal radiation of solid and liquid zirconium carbide emitted by a heated surface per unit area of the sample in the finite frequency range $0.333\,\mathrm{PHz} \le \nu \le 0.545\,\mathrm{PHz}$ at the eutectic temperature 3155 K.

| **Quantity** | **Zirconium Carbide:** <br> *Solid Phase* | **Zirconium Carbide:** <br> *Liquid Phase* | **Gaps** Between *Liquid and Solid Phases* |
|---|---|---|---|
| $I_{\text{total}}(\nu_1, \nu_2, T)$ $[\text{J}]$ | $4.102\times10^{-9}$ | $5.102\times10^{-9}$ | $1\times10^{-9}$ |
| $I^{\text{SB}}_{\text{total}}(\nu_1, \nu_2, T)$ $[\text{W}]$ | $7.379\times10^{2}$ | $9.178\times10^{2}$ | $1.799\times10^{2}$ |
| $A_{\text{total}}$ $[\text{J}]$ | $-1.367\times10^{-9}$ | $-1.701\times10^{-9}$ | $-0.334\times10^{-9}$ |
| $S_{\text{total}}$ $[\text{J K}^{-1}]$ | $1.733\times10^{-12}$ | $2.156\times10^{-12}$ | $0.423\times10^{-12}$ |
| $P_{\text{total}}$ $[\text{J}]$ | $1.367\times10^{-9}$ | $1.701\times10^{-9}$ | $0.334\times10^{-9}$ |
| $C_{V\,\text{total}}$ $[\text{J K}^{-1}]$ | $5.201\times10^{-12}$ | $6.468\times10^{-12}$ | $1.267\times10^{-12}$ |
| $N_{\text{total}}$ | $3.486\times10^{10}$ | $4.336\times10^{10}$ | $0.850\times10^{10}$ |

**Table 2** Calculated values of the thermal radiative and thermodynamic functions of thermal radiation of solid and liquid zirconium carbide emitted by a heated surface area $S$ of the sample in the finite frequency range $0.333\,\text{PHz} \le \nu \le 0.545\,\text{PHz}$ at the eutectic temperature 3155 K.